# Does rotation generate a massive string ?

H Culetu[1]

[1]Ovidius University, Dept. of Physics, B-dul Mamaia 124, 8700 Constanta, Romania

E-mail : hculetu@yahoo.com

**Abstract.** The properties of a stationary massive string endowed with intrinsic angular momentum are investigated. The spacetime is generated by an "improper" time translation combined with uniform rotation. The mass per unit length of the string is proportional to the angular velocity ω. The spacetime is Minkowskian geometrically but the topology is nontrivial thanks to the event horizon located on the surface r = 0 (similar with Rindler's spacetime) and the deficit angle generated by rotation.
The Sagnac time delay is calculated. It proves to be nonvanishing even when ω = 0 due to the intrinsic spin of the string.

The physical interpretation of the parameters from the vacuum solution of Einstein's equations with cylindrical symmetry is still a matter of discussions [1,2]. Some parameters are related to topological defects which have no contribution to the Riemann tensor (the geometry is Minkowskian locally but the topology is nontrivial).
 The spacetime outside an axially symmetric spinning source leads to a topological frame dragging effect [3] associated with some metric parameters responsible for the angular momentum of the string and the deficit angle produced by its mass per unit length [4,5]. The Deser – Jackiw [4] line element is Minkowskian because the presence of the string is hidden. It could be "uncovered" by an "improper" coordinate transformation which mixes the time- and angular- coordinates (the new time variable has a jump whenever the string is circumvented).
 Clement [6] showed that an "illegitimate" coordinate transformation (a Lorent boost in the φ – t plane), combined with a spatial rotation, generates a spinning solution of the static BTZ line element [7]. The Sagnac time delay [8] was computed for matter or light beams counter-propagating on a round trip in a rotating frame. Ruggiero found that the phase shift is nonvanishing even when the observer is not rotating, because of the angular momentum of the source.
 Borrow and Dabrowski [9] studied the case of stringy Godel universes where the velocity of rotation of the universe is related to the inverse string tension by $\Omega^2 = 1/\alpha$'. However, the Riemann tensor in their spacetime is nonvanishing.

We study in this paper the case of a massive infinitely thin string generated by uniform rotation with frequency ω. The string is taken to be "infinitely thin" (or "linelike") in the same manner as an electron, which possesses mass, charge and spin, is considered to be "pointlike" (from the viewpoint of special relativity). Therefore, the matching conditions used in the case of thin shells by Crisostomo and Olea [10] play no role here.

We conjecture the mass μ per unit length of the string is proportional to ω and depends on the Planck constant $\hbar$ through Planck's mass $m_P$. The spacetime has an event horizon located on the surface r = 0 but the surface gravity is imaginary since the horizon is within the time machine region [11,12]. In addition, the Sagnac time delay is computed, depending both on ω and the intrinsic angular momentum of the string.

We use from now on geometrical units with $G = c = \hbar = 1$ and the signature +2.

Let us consider the "improper" coordinate transformation

$$\bar{t} = t + b\varphi, \quad \bar{\varphi} = \omega t + (1+b\omega)\varphi, \quad \bar{r} = r, \quad \bar{z} = z \tag{1}$$

upon the Minkowski line element

$$ds^2 = -d\bar{t}^2 + d\bar{r}^2 + d\bar{z}^2 + \bar{r}^2 d\bar{\varphi}^2 \tag{2}$$

written in cylindrical coordinates $(\bar{t}, \bar{r}, \bar{z}, \bar{\varphi})$. ω < 0 and b are constant with dimensions of frequency and length, respectively.

The transformation (1) alters the metric (2) such that in the new coordinates (t, r, θ, φ) we have

$$ds^2 = -(1-\omega^2 r^2)dt^2 + dr^2 + dz^2 + 2[\omega(1+b\omega)r^2 - b]d\varphi dt + [(1+b\omega)^2 r^2 - b^2]d\varphi^2 \tag{3}$$

The geometry (3) is, of course, flat, but only locally. The situation is similar with that of the Rindler geometry [13] which is flat locally but not globally (the coordinates cover only a part of the Minkowski spacetime because of the event horizon).

We see that the case b = 0 leads to the geometry in a rotating frame of a disk with the constant angular velocity –ω, the so called "Born metric" [14]. However, when ω = 0 is put in (3), the geometry is that of a spinning massless string [4, 11, 15]. We shall take b of the order of Planck's length $l_P = 10^{-33}$ cm. It corresponds to J = (1/4) $l_P$ in the Deser – Jackiw paper. In addition, it could be checked that we reach (3) when Herrera's and Santos' [3] parameters are chosen such that *n = 1, a = 1 and c = - ω*.

We conjecture that the geometry (3) represents the line element viewed by a rotating observer in Minkowski spacetime. By means of the Deser – Jackiw time translation (1) we "uncover" the spinning string which is hidden in Born's coordinates. While b is related to the intrinsic spin per unit length, the product bω may be written as

$$-b\omega = \frac{-b}{m_P} \frac{m_P \omega}{c} \equiv \frac{G}{c^2}\mu \tag{4}$$

where μ = $m_P$ |ω| /c is the mass per unit length of the string (we used all the fundamental constants in the above equation, for clarity).

In terms of the "light cylinder" radius l = 1/ω, we have μ = $m_P$/l. With, for example, ω = 300 s$^{-1}$, one obtains μ = $10^{-13}$ g/cm. For an observer located at r >> b with respect to the axis of rotation, $g_{\varphi\varphi}$ from (3) becomes

$$g_{\varphi\varphi} \approx (1-\mu)^2 r^2 \tag{5}$$

i.e. the range of φ is diminished to 2π (1-μ) [4] (a deficit angle appeared). We have, of course, a similar deficit angle even with ω = 0. That case gives

$$g_{\varphi\varphi} = (1 - \frac{b^2}{r^2})r^2 \tag{6}$$

However, the defect in the angular range is now r – dependent.

To find the location of the timelike Killing horizon H of the rotating observer, we must apply the formula [11]

$$\hat{g}_{tt} \equiv g_{tt} - \frac{g_{t\varphi}^2}{g_{\varphi\varphi}} = 0, \tag{7}$$

since the metric (3) is not diagonal. The equation (7) yields $r_H = 0$, irrespective of the values of $\omega$ and $b$. The region with $g_{\varphi\varphi} < 0$ or $r < b/|1+b\omega|$ is the time-machine region. Its boundary is "the velocity of light surface".

The expression for the surface gravity $\kappa$ of the horizon appears as

$$\kappa^2 = \nabla_\alpha L \nabla^\alpha L \tag{8}$$

where

$$L^2 = -g_{tt} - 2\Omega_H g_{t\varphi} - \Omega_H^2 g_{\varphi\varphi} \tag{9}$$

and the angular velocity $\Omega_H$ of the horizon (with respect to a nonrotating observer at infinity) is given by

$$\Omega_H = \left(-\frac{g_{t\varphi}}{g_{\varphi\varphi}}\right)\Big|_H = -\frac{1}{b} \tag{10}$$

With $\Omega_H$ from the above equation, we obtain $L^2 = -r^2/b^2$, which gives an imaginary $\kappa$, with $|\kappa| = 1/b$. An imaginary surface gravity leads to an imaginary horizon temperature $T = \kappa/2\pi$. The reason may be related to the fact that the horizon is located inside the time-machine region.

Let us study now the Sagnac effect in the spacetime (3). We shall use two counter-propagating light beams on a circular trajectory (by means of a system of mirrors). We assume the source of light is at rest in the rotating frame at $r = r_0$, $z = 0$. With $ds^2 = 0$ in (3), we get, for the velocity $d\varphi/dt$, with respect to the asymptotically rest frame [16]

$$\left(\frac{d\varphi}{dt}\right)^2 + \frac{2[(1+b\omega)\omega r_0^2 - b]}{(1+b\omega)^2 r_0^2 - b^2}\frac{d\varphi}{dt} - \frac{1-\omega^2 r_0^2}{(1+b\omega)^2 r_0^2 - b^2} = 0 \tag{11}$$

Solving the above equation for $d\varphi/dt$, one obtains

$$\Omega_\pm \equiv \left(\frac{d\varphi}{dt}\right)_\pm = -\frac{(1+b\omega)\omega r_0^2 - b}{(1+b\omega)^2 r_0^2 - b^2} \pm \frac{r_0}{(1+b\omega)^2 r_0^2 - b^2} \tag{12}$$

whence, for the two counter-rotating beams

$$\Omega_+ = \frac{1-\omega r_0}{r_0 - b + b\omega r_0}, \qquad \Omega_- = -\frac{1+\omega r_0}{r_0 + b + b\omega r_0} \tag{13}$$

Let us notice that the particular case $b = 0$ in (12) gives $\Omega_\pm = -\omega \pm (1/r_0)$, a result obtained by Soler [17]. In addition, the case $\omega = 0$ leads to $1/(\pm r_0 - b)$ [18]. We observe that $\Omega_+ > 0$ and $\Omega_- < 0$ outside the time machine region ($r_0 > b/(1-b|\omega|)$) and inside the light cylinder ($r_0 < 1/|\omega|$) (we consider that $|\omega|$ is less that its Planck value $1/b$).

Solving for $t_+$ and $t_-$ from (12), we have for the Sagnac time delay [8, 12]

$$\Delta t \equiv t_+ - t_- = \frac{4\pi\omega r_0^2}{1-\omega^2 r_0^2} - 4\pi b. \tag{14}$$

We observe that $\Delta t$ depends separately on the two parameters $\omega$ and $b$. When $\omega = 0$, $\Delta t$ corresponds to the value obtained in [18], but when $b = 0$, Soler's result emerges. On the "light cylinder" the time delay becomes infinite.

To summarize, several properties of a massive string endowed with angular momentum have been discussed in this paper. The spacetime is Minkowskian only locally but the topology is nontrivial due to a time jump and a defect in the angular range (deficit angle) introduced by rotation with constant frequency.

We supposed the massive string was generated by the agent who rotates the physical system. The horizon is located on the axis of rotation and the modulus of the surface gravity is $1/b$, where $b$ is the Planck length. The mass per unit length of the string is proportional to ω and depends on $\hbar$.

A Sagnac phase shift will appear between two counter-propagating light beams due to the intrinsic angular momentum of the string (proportional to $b$) and the angular velocity of the rotating observer.

**Acknowledgements**

I would like to thank the Organizers of ERE 06 Workshop for inviting me to attend this very stimulating meeting and for their warm hospitality during my stay at Palma de Mallorca, Spain, where this paper has been presented.